\documentclass[%
 apl,%
 amsmath,%
 amssymb,%
 amsfont,%
 draft
 preprint,%
]{revtex4-1}

\usepackage{graphicx}
\usepackage{bm}
\begin{document}

\title{ Atomistic study of electronic structure of PbSe nanowires }%
\author{Abhijeet Paul}
\email{abhijeet.rama@gmail.com}
\author{Gerhard Klimeck}

\affiliation{School of Electrical and Computer Engineering, Network for Computational Nanotechnology, Purdue University,%
 West Lafayette, Indiana, USA, 47907.}

\date{\today}

\begin{abstract}

Lead Selenide (PbSe) is an attractive `IV-VI' semiconductor material to design optical sensors, lasers and thermoelectric devices. Improved fabrication of PbSe nanowires (NWs) enables the utilization of low dimensional quantum effects. The effect of cross-section size (W) and channel orientation on the bandstructure of PbSe NWs is studied using an 18 band $sp^3d^5$ tight-binding theory. The bandgap increases almost with the inverse of the W for all the orientations indicating a weak symmetry dependence. [111] and [110] NWs show higher ballistic conductance for the conduction and valence band compared to [100] NWs due to the significant splitting of the projected L-valleys in [100] NWs.

%%%Lead Selenide (PbSe) is an attractive material to design optical sensors, lasers and low temperature thermoelectric materials. Recent technological advances enabled the fabrication of PbSe nanowires (NWs) thus allowing to utilize quantum effects at low dimensionality. The effect of cross-section size and channel orientation on PbSe NWs is studied using a relativistic 18 band $sp^3d^5$ tight-binding theory. Conduction and valence band L-valleys show significant valley splitting in [100] orientation compared to [110] and [111]. The bandgap increases almost as the inverse of the NW cross-section size for all the orientations. [111] and [110] NWs show higher ballistic conductance (at 10K) for conduction and valence band compared to [100] NWs.     

\end{abstract}


\maketitle %%
%\section{Introduction} \label{sec:I}
\textit{Appealing PbSe bulk properties:} Lead selenide (PbSe) is a narrow, direct band gap semi-conductor material ($\sim$ $Eg^{bulk}$ = 0.16eV at 4K \cite{PbSe_data,Lent_TB})  with useful electrical, optical and lattice properties \cite{PbSe_optic,PbSe_bulk_thermo,MEgen_1}. It is used extensively in optical devices \cite{Kang_and_wise,PbSe_optic}, lasers \cite{PbSe_laser,PbSe_laser_2} and thermoelectric devices \cite{PbSe_NW_fab_3,PbSe_bulk_thermo,PbSe_NW_thermo}. The large Bohr exciton radius of about 46nm in PbSe makes it a suitable system to study quantum confinement effects on electrons and holes \cite{Kang_and_wise,PbSe_NW_fab_1,PbSe_NW_fab4,PbSe_qdot}. Recent progress in multiple exciton generation (MEG) in PbSe with higher optical efficiency has renewed interest in the optical properties of PbSe \cite{MEgen_1}. Extremely low thermal conductivity of PbSe ($\sim$ 2 W/m-K in bulk \cite{PbSe_low_kappa} to $\sim$ 0.8 W/m-K in NWs \cite{PbSe_NW_low_kappa})  also makes it a suitable thermoelectric material \cite{PbSe_bulk_thermo,PbSe_NW_thermo}. PbSe can become a preferable material over Lead Telluride (PbTe) due to the higher availability of selenium (Se) compared to tellurium (Te) \cite{PbSe_singh_DFT}. 

\textit{Nanostructures PbSe :} One dimensional nanostructures of PbSe like nanowires (NW) and nanorods (NR) combine the interesting bulk material properties as well as the quantum confinement effects which can lead to better thermoelectric \cite{PbSe_NW_thermo,PbSe_low_kappa,PbSe_NW_low_kappa} and optical \cite{PbSe_optic} devices. The analysis of the physical properties in PbSe NWs will require proper understanding of the electronic structure, which is the theme of the present paper.

%Lead chalcogenides are very interesting semiconductor materials due to their optical and electronic properties. These are the low bandgap materials with symmetric hole and electron mass and high carrier mobility. Also these materials have very large excitonic radius of about 46nm which allows for good carrier confinement.
 %The extremely low thermal conductivity of these materials also make them viable low temperature thermoelectric materials. These interesting applications for the lead salts make them very important semiconductor material.

%Lead Selenide (PbSe) one of the lead salts is an interesting material for optical, magnetic as well as thermoelectric applications. 
%The PbSe nanowires are quite useful for making the next generation optical and thermoelectric devices \cite{PbSe_bulk_thermo,PbSe_QW_thermo,PbSe_NW_thermo}.

\textit{PbSe NW growth technologies:}
With recent advances in the growth and process technology, the fabrication of PbSe NWs have become very efficient and controlled. PbSe NWs are developed using a variety of methods like, chemical vapor transport (CVT) method \cite{PbSe_NW_fab_3}, oriented nano-particle attachment \cite{PbSe_NW_fab_1}, electro-deposition without catalyst \cite{PbSe_NW_fab4} and with catalyst \cite{PbSe_NW_fab_5}, hyper-branching \cite{PbSe_NW_fab_2}, growing PbSe structures on phosphate glass \cite{PbSe_qdot_fab}, etc. These methods enable PbSe NWs fabrication with a variety of growth directions and surfaces.

\begin{figure}[!t]
	\centering
		\includegraphics[width=3.4in,height=1.8in]{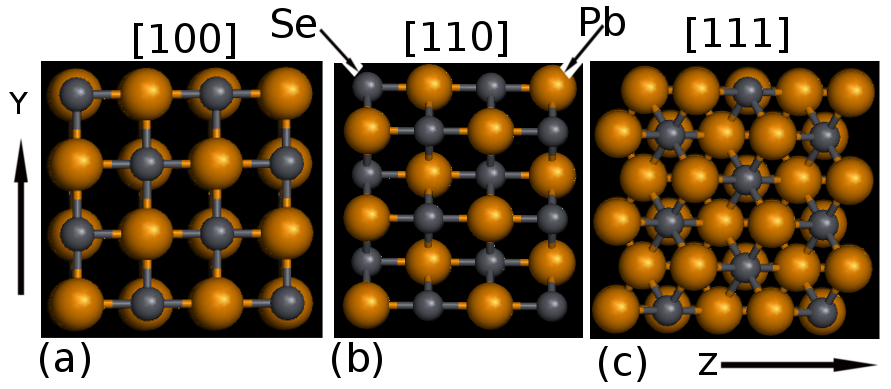}
	\caption{Projected PbSe unitcell with wire axis orientations (X) along (a) [100], (b) [110] and (c) [111]. The cross-section size is of 2.5nm (Y) $\times$ 2.5nm (Z). Lead (Pb) and selenium (Se) atoms are shown in gold and grey color respectively.}
	\label{fig:ucell_nws}
\end{figure}

\textit{PbSe crystal structure:} Bulk PbSe has a stable rocksalt structure with a co-ordination number of six at room temperature (300K) and normal atmospheric pressure \cite{PbSe_singh_DFT,PbSe_data}.  The lattice constant is 0.6121 nm \cite{PbSe_data} at T = 4K, which is also utilized in our bandstructure calculations. The PbSe NWs are constructed using the same bulk structure with three different wire axis orientations of [100], [110] and [111] (Fig. \ref{fig:ucell_nws}). 

\textit{Electronic Structure calculation:} The main methods for analyzing the lead salt structures are (1) continuum method like 4/8-band k$\cdot$p \cite{Kang_and_wise}, (2) semi-empirical atomistic methods like tight-binding (TB)\cite{delure_TB,Lent_TB}, pseudo-potential approaches \cite{PbSe_EPM_2} or (3) first principles methods \cite{PbSe_qdot,PbSe_DFT,PbSe_DFT_1,PbSe_singh_DFT}. The 4/8-band k$\cdot$p method captures some quantum confinement effects, however, it fails to capture the interaction of various valleys present along the \textit{L-K} and \textit{L-X} directions, the band anisotropy and the correct frequency dependent dielectric function  \cite{PbSe_qdot,PbSe_EPM_2,delure_TB}. The first principle calculations are highly accurate but the demand for computational power is very high and these methods are limited to solving only small structures with few thousand atoms \cite{Klimeck_TB,delure_TB}. The semi-empirical methods are versatile in terms of the involved physics and can handle a large number of atoms ($\ge$ 10 million atoms \cite{Klimeck_TB}) making them suitable for electronic structure and transport calculation in realistic device structures. However, an integral part of these methods is the requirement of correct semi-empirical parameter sets to properly represent the electronic structure properties like bandgaps, effective masses, wave-function symmetry, etc. In this work we utilize a semi-empirical atomistic TB method based on $sp^3d^5$ formulation with spin orbit coupling (SOC) optimized for bulk Pb salts by \textit{Lent et al} \cite{Lent_TB} to calculate the electronic structure in NWs. The inclusion of SOC is important in PbSe since both CB and VB have strong p-contributions from Pb and Se atoms, respectively \cite{PbSe_qdot,delure_TB}. 

\textit{Surface atoms:} The TB parameters developed for bulk Pb salts \cite{Lent_TB} are also applied to nanostructures with finite boundaries. The surface atoms are not passivated for Pb salt nanostructures. It has been previously shown that surfaces do not introduce states in the energy gaps of the bulk band structure, even though they are not passivated \cite{delure_TB} and the same is obtained in the present study. The surface states are mainly p-like for Pb salts which are strongly coupled with orbitals of atoms at the interior unlike the zinc-blende semiconductors where atoms are described by hybrid $sp^3$ orbitals which remain uncoupled at surfaces, forming dangling bonds \cite{Lee_sp3}. The lack of surface passivation has also been pointed out by the first principle calculations in stoichiometric Pb salt nanostrutures \cite{PbSe_DFT_1,PbSe_qdot}.

%% This is in contrast with zincblende semiconductors 
%% For the calculation of bandstructure within TB model we also need to passivate the surface atoms in order to remove the band-gap states. One important point in the lead salts is that passivation is not necessary for nanostructures especially if the structures are stoichiometric as pointed out it Ref.[]. There have been other TB implementations for thin films where passivation has not been done and no band-gap states are observed. In our calculations too we did not observe any band-gap states without passivation.

\begin{figure}[!t]
	\centering
		\includegraphics[width=3.2in,height=2.3in]{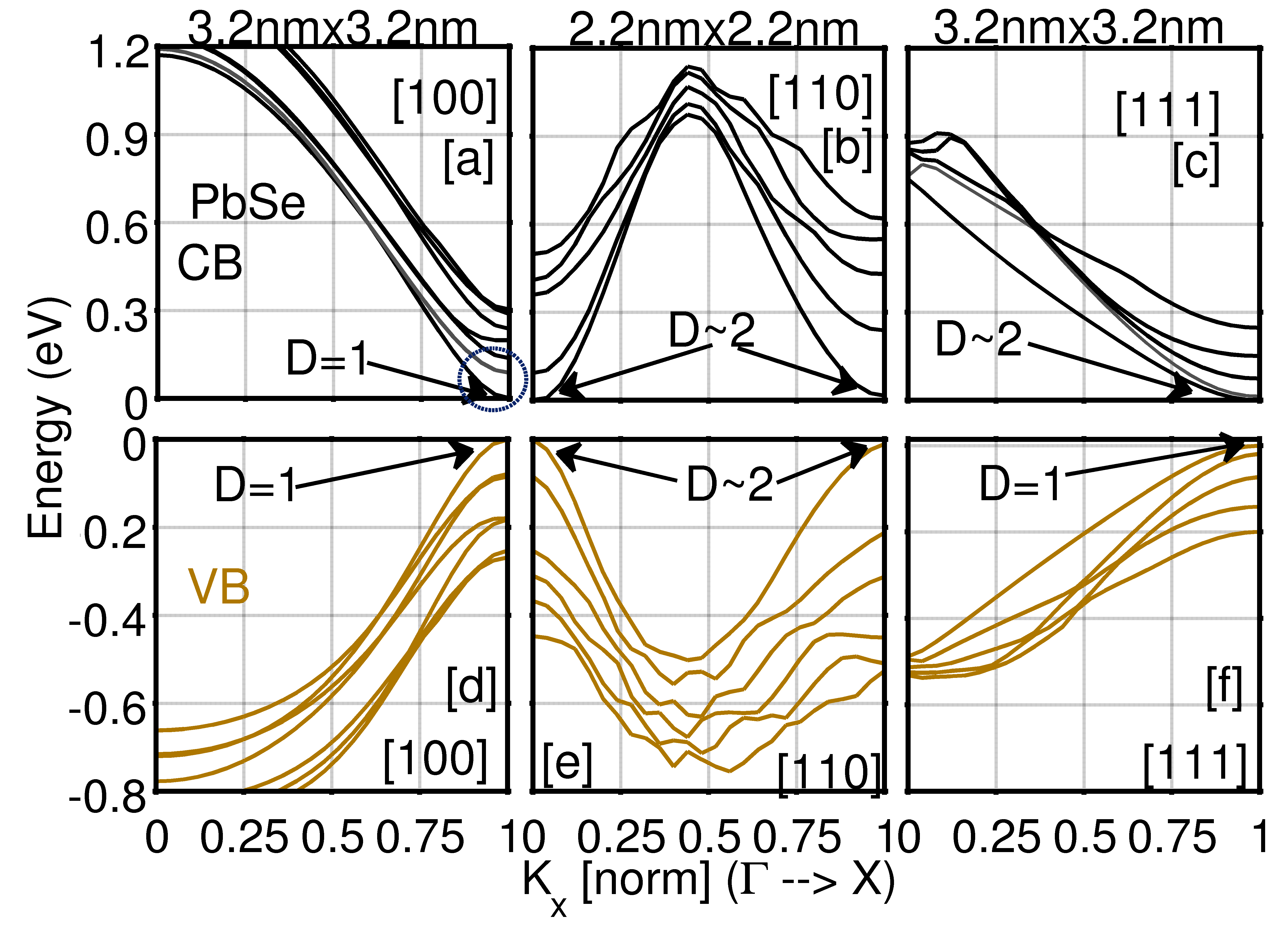}
	\caption{Bandstructure of PbSe NWs for (a) [100], (b) [110] and (c) [111] oriented channels. The CB minima and VB maxima are normalized to zero for simplified band comparison. Only the first 5-10 sub-bands are shown.}
	\label{fig:EK_all_nw}
\end{figure}

\textit{Electronic Structure:} The bulk effective masses calculated using the TB parametrization in Ref. \cite{Lent_TB} is obtained for (i) electrons as $m_{e}^{\parallel}/m_{e}^{\perp}$ = $0.087(0.07)/0.036(0.04)$ = 2.4~(1.85) and (ii) holes as $m_{h}^{\parallel}/m_{h}^{\perp}$ = $0.094(0.068)/0.031(0.34)$ = 2.9~(2.0). The values in the parenthesis are from Ref.\cite{PbSe_data}. The TB parametrization in Ref. \cite{Lent_TB}($TB_{A}$) is preferred over the parametrization in Ref. \cite{delure_TB} ($TB_{B}$) since the $TB_{A}$ model captures the mass anisotropy quite well, an important requirement for TB models as pointed in Ref. \cite{PbSe_EPM_2}. 

The bandstructure of conduction (CB) and valence band (VB) for PbSe nanowires are shown in Fig. \ref{fig:EK_all_nw} for 3 different wire orientations. For all the wires the conduction band minima (CBM) and valence band maxima (VBM) are normalized to zero to enable a better comparison of the valleys. Few important points to observe are, (i) bulk L valleys are projected at the Brillioun zone (BZ) edge at X for all orientations (Fig. \ref{fig:EK_all_nw}). The [110] wire has an additional projected valley at the $\Gamma$ position (Fig.\ref{fig:EK_all_nw} b, e), (ii) [100] NW show a degeneracy (D) of 1 for both CB and VB due to strong valley splitting, however [110] and [111] NWs show a D of nearly 2 (extremely small valley splitting) for the positive `k' states.  

% gives a degenracy of 1 for these two orientations. However, for [110] case the degenracy is 2 since one of the bulk L valley is also projected at the zone center (at $\Gamma$ point). 

%Thus [110] wire will provide a higher conductance compared to the [100] and [111] case for the same Fermi-level position.

\begin{figure}[!b]
	\centering
		\includegraphics[width=3.4in,height=1.75in]{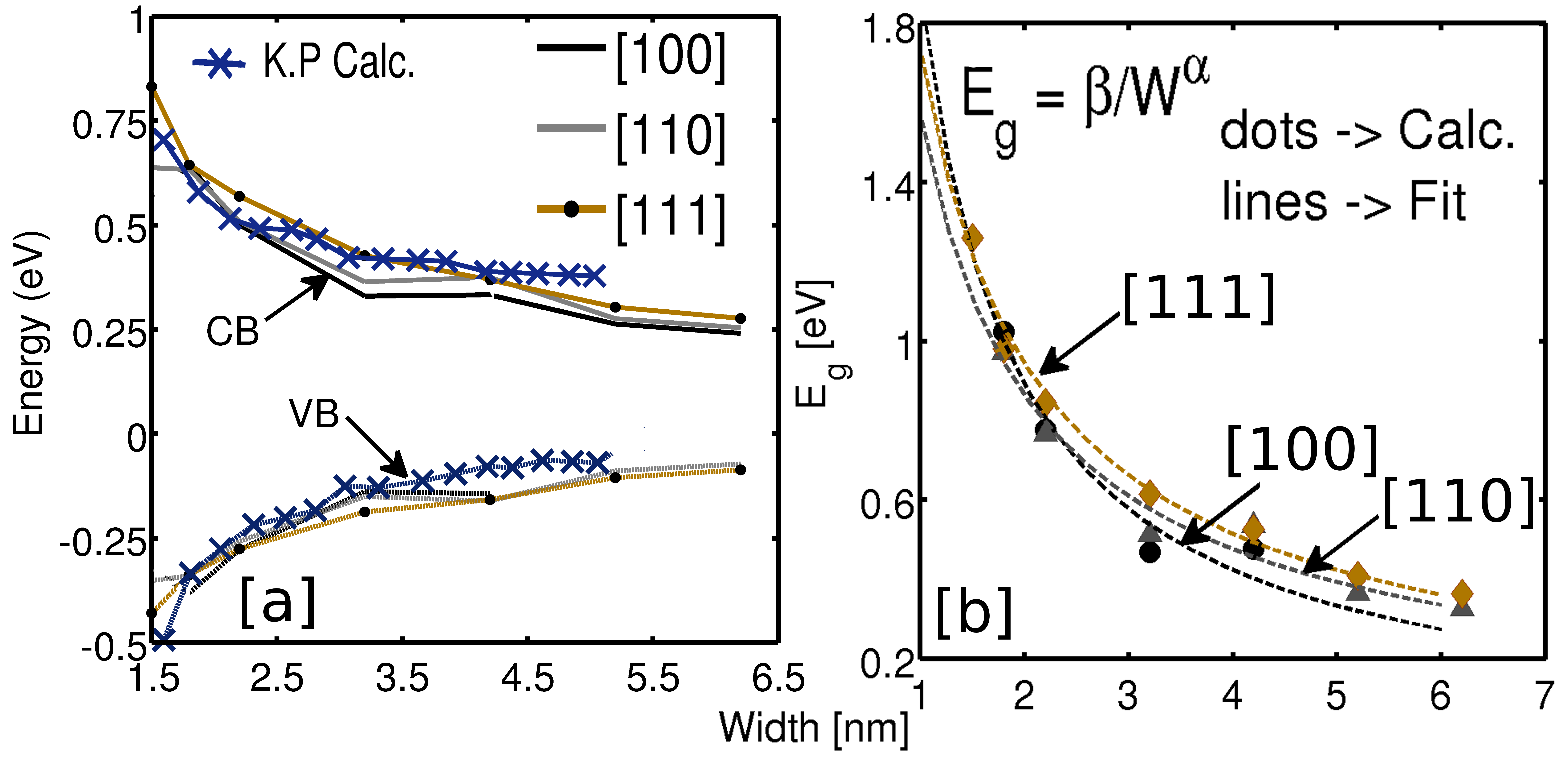}
				%%{Band_Edges_PbSe_NW_with_size_all_oreint_new.png}
	\caption{(a) Bandedges for square PbSe nanowires with [100], [110], and [111] channel orientations. Bandedge result using 4-band k$\cdot$p calculation for cylindrical PbSe nanowires from Ref. \cite{PbSe_NW_kp}. (b) Bandgap variation for all the nanowires. Dots represent calculated TB values and lines represent analytical fitting with cross-section size.}
	\label{fig:band_edges_nw}
\end{figure}

\textit{Bandgap and bandedge variation:} The CBM and VBM variation with cross-section size (W) and orientation are shown in Fig. \ref{fig:band_edges_nw} a. All the orientations show quite similar band-edge variation with W. As the cross-section size decreases the geometrical confinement increases which pushes the CBM (VBM) higher (lower) in energy. The variation in the bandedges with W also compare surprisingly well with a previous 4-band k$\cdot$p calculation done for [111] cylindrical PbSe NWs \cite{PbSe_NW_kp}. The bandgap variation with W in PbSe NWs can be fitted to the following analytical expression,

\begin{equation}
Eg = \beta/(W)^{\alpha},
\end{equation}
where $\alpha$ represents the power law dependence on W.

For [111] NWs the value of $\beta,(\alpha)$ is 1.726 (0.8734). For [110] NWs these values are ($\beta,(\alpha)$) 1.564 (0.8592) and for [100] NWs these values are ($\beta,(\alpha)$) 1.87(1.072).  The bandgap values (Fig. \ref{fig:band_edges_nw} b) roughly show an inverse relation with W for all the NW orientations which is very different from the prediction of effective mass approximation (Eg $\propto$ $W^{-2}$). Similar results for Eg have been obtained by other independent calculations carried out in PbSe nanostructures using first principle calculations \cite{PbSe_qdot} as well as TB calculations \cite{delure_TB}. This justifies the application of TB electronic structure calculation which correctly captures the quantum confinement effects in ultra-scaled PbSe NWs.

\begin{figure}[!t]
	\centering
		\includegraphics[width=3.4in,height=1.8in]{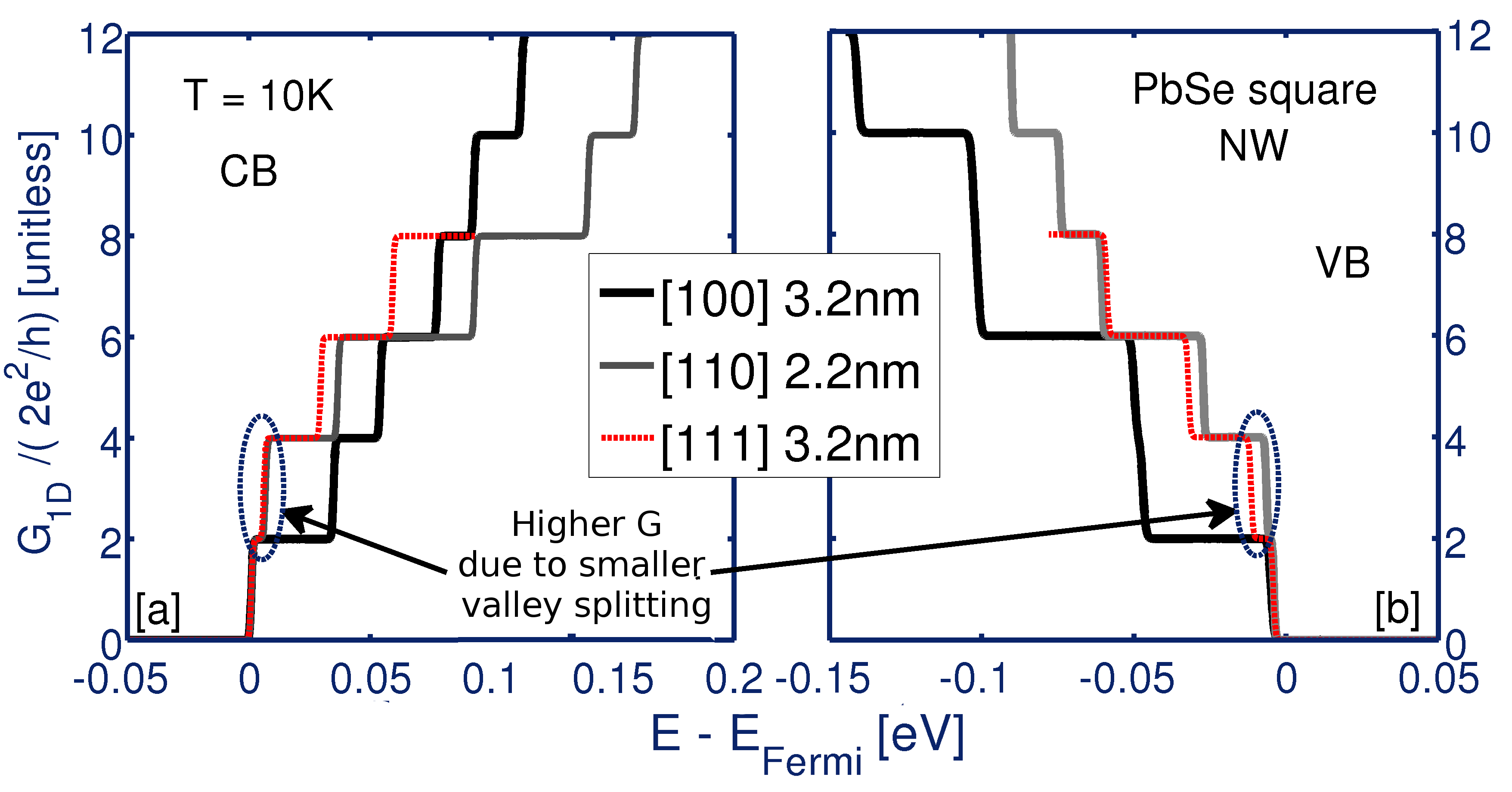}
	\caption{Ballistic conductance in PbSe NWs for (a) CB and (b) VB for 3 different wire orientations at T = 10K. The temperature comes from the Fermi-Dirac distribution of the carriers.}
	\label{fig:TE_all_nw}
\end{figure}

\textit{Ballistic conductance:} Transport properties of PbSe NWs are revealed by the electronic conductance which is calculated using Landauer's formula \cite{Land}. Figure \ref{fig:TE_all_nw} shows the normalized 1D ballistic conductance for electrons in the PbSe NWs for 3 different orientations. [110] and [111] oriented wires show higher G value for both the CB and the VB compared to [100] NWs. A larger valley splitting in the CB and the VB in [100] NWs decrease the conductance compared to the other two orientations. The normalized conductance increases in steps of 2 for [100] NWs for the CB but not for the VB (Fig. \ref{fig:TE_all_nw}) which shows that the CB and the VB are not entirely symmetric in energy, a result similar to the one given in Ref. \cite{PbSe_EPM_2}. Thus, transport characteristics show the influence of geometrical confinement and channel orientation. 

\textit{Summary and Outlook:} PbSe NWs have tremendous potential of becoming the next generation thermoelectric and optical devices. The proper understanding of the physical properties of these ultra-scaled PbSe nanowires will depend strongly on the correct electronic structure calculation. We have presented the application of semi-empirical tight-binding theory to these NWs to understand the variation of conduction and valence bands and the position of the important energy valleys. The variation of the bandgap and the bandedges with NW cross-section size is strongly influenced by the cross-section size and channel orientation. Simple effective mass theory cannot predict the bandgap variation. The amount of valley splitting is strongly dependent on the type of geometrical confinement as reflected in the ballistic conductance of the PbSe NWs. [110] NWs provide the maximum ballistic conductance for both the CB and VB. The tight-binding analysis of electronic structure opens door to explore the optical and thermoelectric effects in PbSe NWs.

%The authors acknowledge financial support from MSD Focus Center, under the Focus Center Research Program (FCRP), a Semiconductor Research Corporation (SRC) entity, Nanoelectronics Research Initiative (NRI) through the Midwest Institute for Nanoelectronics Discovery (MIND), NSF (Grant No. OCI-0749140) and Purdue University. Computational support from nanoHUB.org, an NCN operated and NSF (Grant No. EEC-0228390) funded project is also gratefully acknowledged.

Financial support from MSD Focus Center, under SRC, Nanoelectronics Research Initiative through MIND, NSF (Grant No. OCI-0749140) and Purdue University is acknowledged. Computational support from nanoHUB.org, an NCN operated and NSF (Grant No. EEC-0228390) funded project is also gratefully acknowledged.

%\nocite{*}
%\bibliographystyle{natbib}
\bibliography{refs}

\end{document}